\documentclass[12pt]{article}

\usepackage[letterpaper,margin=1in]{geometry}
\usepackage{setspace}
\usepackage{titlesec}

\renewcommand{\thesection}{\Roman{section}}
\renewcommand{\thesubsection}{\Alph{subsection}}

\titleformat{\section}
  {\normalfont\normalsize\centering}
  {\normalfont\thesection.}
  {0.5em}
  {\bfseries}

\titleformat{\subsection}
  {\normalfont\normalsize\centering}
  {\thesubsection.}
  {0.5em}
  {}

\usepackage{libertinus}
\usepackage{amsmath}
\usepackage{amssymb}
\usepackage{bm}
\usepackage{pdflscape}

\usepackage{graphicx}
\usepackage{booktabs}
\usepackage{threeparttable}
\usepackage{tabularx}
\usepackage{array}
\usepackage{multirow}
\usepackage{longtable}
\usepackage{subcaption}
\usepackage{caption}
\usepackage{float}
\usepackage{rotating}

\usepackage[round,authoryear]{natbib}
\usepackage{xcolor}
\definecolor{darkblue}{rgb}{0.0,0.0,0.60}
\usepackage{tikz}
\usetikzlibrary{arrows.meta}
\usepackage{pgfplots}
\pgfplotsset{compat=1.18}

\PassOptionsToPackage{hyphens}{url}
\usepackage{hyperref}
\hypersetup{
    colorlinks=true,
    urlcolor=darkblue,
    linkcolor=darkblue,
    citecolor=darkblue
}
\usetikzlibrary{arrows.meta,decorations.pathreplacing}

\usepackage{microtype}
\usepackage{enumitem}

\begin{document}


\title{
\textbf{Outsourced Cryptocurrency Wash Trading}
\thanks{
I am grateful to Lin Peng, William Cong, and seminar participants at the
Baruch--SWUFE seminar for helpful comments.
A previous version of this paper was circulated under the title
``How Wash Traders Exploit Market Conditions in Cryptocurrency Markets.''
}
}

\author{
Hunter Ng\\
\small Baruch College, City University of New York
}

\date{}

\maketitle


\begin{abstract}
\noindent
I show that outsourced cryptocurrency wash trading is shaped by the objective assigned to the outside firm. When hired to manufacture volume, wash traders buy and sell nearly equal amounts, repeatedly turning over the same inventory. When also tasked with selling issuer holdings, trading becomes markedly more one-sided, with sales exceeding purchases. Using documented enforcement cases and transaction records, I trace these differences in net selling, two-way trading, and inventory replenishment across controlled wallets. The results show that similar reported trading volume can mask fundamentally different economic activity as contractual objectives shape whether manipulation recycles inventory or liquidates issuer holdings.
\end{abstract}

\medskip


\medskip


\clearpage

\section{Introduction}
A growing literature has studied wash trading in cryptocurrency and digital-asset
markets,\footnote{The SEC defines a wash sale as a transaction in which beneficial ownership remains unchanged, and a manipulative transaction that can create the appearance of market activity
\citep{sec_wash_sales}}
focusing on its detection and prevalence
\citep{cong_crypto_2023,aloosh_direct_2024,wang_wash_2025}. Related to this literature, recent SEC and DOJ enforcement actions from 2022 to 2024 provide the first detailed public record of wash trading organized as an outsourced commercial service by specialized firms \citep{sec_market_makers_2024,doj_token_mirrors_2024}. This differs markedly from an earlier documented case of Mt.~Gox, in which leaked internal records identify wash trading through matching trader IDs
\citep{gandal_price_2018,aloosh_direct_2024}. Outsourcing introduces a distinct incentive problem because the objectives assigned to the outside firm can directly shape how illegal market manipulation is executed. Understanding how these incentives affect manipulative trading is important for detecting misconduct and designing effective enforcement \citep{becker_crime_1968,zitzewitz_forensic_2012}.

I find that external wash traders hired mainly to create trading volume tend to buy and sell similar amounts of tokens, generating substantial apparent activity while transferring little inventory to outside investors. When the issuer also instructs the trader to sell its tokens, trading becomes more one-sided, with sales exceeding purchases even as substantial trading activity continues. Reported volume alone can conceal important differences in the economic substance of manipulation. The objective assigned to the outside firm determines whether wash trading mainly recycles the same tokens or also facilitates persistent sales of issuer inventory, which is changes the detection of wash trading \citep{cong_crypto_2023}.

To interpret these findings, I turn to the enforcement records. The 2022-2024 SEC and DOJ
cases provide three advantages over existing settings. First, unlike most existing
studies, which infer wash trading from statistical patterns in observed transactions,
these cases provide documentary evidence of wash trading. Such evidence has
traditionally been difficult to obtain because wash trading can be difficult to
distinguish from legitimate trading based on transaction data alone.\footnote{In the Mt.~Gox case, leaked internal account
records provided direct evidence of wash trading through matching trader IDs. The recent SEC and DOJ
enforcement actions provide a different source of evidence through legal records.}
The enforcement records identify the parties involved and document the underlying
wash-trading arrangement. Second, the circumstances surrounding the wash trading
are extensively documented. The enforcement records describe the relationships
between the parties and how the wash trading was organized and carried
out. This is different from other settings where the transaction records are available but no other circumstantial evidence is available. Figure~\ref{fig:internal_external_contractibility}
classifies selected documented cryptocurrency wash-trading cases by organization and trading objective. Supplemental Appendix Table~B1 records the full case inventory and related proposed engagements; Figure~\ref{fig:internal_external_contractibility} displays selected entries for visual clarity. Third, the records make the contractual objectives of the wash trader
observable. They describe what the outsourced wash trading firm was hired
to accomplish, including generating trading volume, selling tokens held by the
sponsor, or both.

Using the trading records of these documented wash-trading cases and their documented contracts, I find that the objectives assigned to the outside firm are reflected in how it trades. When the firm is hired mainly to generate volume, purchases and sales closely offset one another, allowing substantial trading activity to be generated with very little net selling. When the firm also receives an issuer-sales instruction, sales rise relative to purchases and net selling accounts for a larger share of gross activity. Trading composition and net selling differ across the two engagements in a manner consistent with their contractual objectives.

I first compare total purchases and sales over each trading episode. In the volume-oriented arrangement, purchases and sales are almost equal, and nearly all observed activity takes the form of two-way trading. In the arrangement that also includes an issuer-sales instruction, sales exceed purchases by substantially more. The volume-oriented firm consequently generates far more trading activity for each token sold on net. These patterns are consistent with the model’s prediction when the issuer-sales target binds and inventory and market-absorption constraints are slack.

I next examine whether this trading composition persists over shorter horizons. A firm may generate balanced purchases and sales over an entire episode through trading that is highly directional within shorter intervals. I therefore divide the CLS trading history into shorter intervals and measure how closely purchases and sales offset within each interval. The volume-oriented arrangement remains highly balanced at six-hour and daily horizons, and its observed balance exceeds the conditional upper bound implied by the reported HYDRO company-account purchases and sales.

Finally, I examine how this balance is generated across the wallets controlled by the outside firm. Individual wallets exhibit substantial directional trading, while these positions offset extensively once the wallets are aggregated under common control. I also find that most tokens sold by the volume-oriented firm are replenished through purchases in the controlled wallet cluster within a day. The trading network therefore generates high activity through offsetting purchases and sales across controlled accounts.

Taken together, these results show differences in trading composition across the two contractual objectives. The volume-generation assignment exhibits balanced, repetitive turnover, while the issuer-sales assignment exhibits substantially more net selling. The key implication is that the same reported trading volume can represent very different economic activity depending on the objective given to the wash trader.


\section{Institutional Details}
\label{sec:institutional}
Wash trading in cryptocurrency markets is often studied using indirect evidence. Researchers typically infer manipulation from statistical irregularities in publicly observed transactions, including unusual trade-size distributions, deviations from Benford's law, abnormal volume patterns, and structural breaks in trading activity. These approaches can identify patterns consistent with wash trading, but these patterns alone cannot conclusively establish that particular
transactions were wash trading or identify the parties controlling the accounts involved. \citet{gandal_price_2018} and \citet{aloosh_direct_2024} provide a rare exception using leaked internal records from Mt.~Gox that identify the traders on both sides of each transaction. The Mt.~Gox data were unusual precisely because the exchange's internal transaction records became public following its collapse, allowing wash trading to be identified directly rather than inferred from trading patterns. The SEC and DOJ enforcement actions announced between 2022 and 2024 provide a new set of independently documented cases. Although there could be many other instances of undocumented wash trading, these cases provide an authoritative basis for identifying the parties involved and the underlying contracts.

\citet{cong_crypto_2023} examine trading records from 29 centralized
cryptocurrency exchanges and develop statistical tests for identifying
artificial trading. They find substantial abnormalities in the trading records
of unregulated exchanges and estimate that wash trading accounts for more than
70 percent of reported volume on these exchanges, on average. They further
show that fabricated volume improves exchange rankings and can temporarily
affect prices. The authors also report that their research was consulted by
the FBI and other U.S. authorities to assist federal investigations
\citep{cong_crypto_2023}.

\citet{lehar_parlour} examine whether decentralized exchange can perform
the economic functions of a conventional market. Combining a model of
liquidity provision with transaction-level evidence from Uniswap, they
document the absence of persistent arbitrage opportunities and identify
conditions under which automated market making outperforms a limit order
market. Their findings concern the efficiency of exchange, which is
distinct from whether observed trading reflects independent investor
demand. A market can execute transactions efficiently even when some
participants trade specifically to create the appearance of activity.
Indeed, SEC enforcement records describe arrangements in which token
promoters hire specialized firms to manufacture trading volume
\citep{sec_market_makers_2024}. In such arrangements, volume could also be determined by outsourced wash traders. This leaves
a gap in the literature in understanding how different objectives are implemented within outsourced wash trading.

\citet{amiram_lyandres_rabetti_2025} study the incentives of cryptocurrency
exchanges to inflate trading volume. They find that greater competition among
exchanges is associated with greater volume inflation. Exchanges receive
short-run benefits from inflated activity but subsequently incur reputational
costs. Together with \citet{cong_crypto_2023}, these findings suggest that
manufactured trading volume should be viewed as an economic objective rather
than simply as a by-product of other manipulative activity.

\citet{aloosh_direct_2024} use internal transaction records from Mt.~Gox
to identify more than 115,000 trades in which the same trader appears on
both sides. They use these trades to evaluate statistical methods for
detecting wash trading. They find support for tests based on Benford's
law, trade-size clustering, lognormal distributions, and structural
breaks, whereas tests based on power-law tails can give misleading
conclusions. Their own index, based on unusually high trade counts and
small median trade sizes, performs as well as, or better than, the
alternative methods they compare for identifying short periods containing
wash trades. Their findings show the value of direct evidence for
evaluating statistical detection methods. Similar to their setting, this paper uses documented cases of wash trading and their documented wash-trading arrangements.

Drawing on the SEC and DOJ enforcement records, I summarize selected documented
cryptocurrency wash-trading cases in
Figure~\ref{fig:internal_external_contractibility}. The different outsourced arrangements raise the question of how contractual objectives relate to the pattern of wash trading. This complements existing research on the conduct and
detection of wash trading and also connects
to a longstanding literature in forensic economics on the incentives underlying illegal
behavior and their implications for enforcement
\citep{becker_crime_1968,polinsky_shavell_enforcement_2000, zitzewitz_forensic_2012}.
\begin{figure}[htbp]
\centering

\begin{tikzpicture}[
    x=1.15cm,
    y=1.05cm,
    axis/.style={
        line width=0.9pt,
        -{Stealth[length=2.8mm,width=2mm]}
    },
    casepoint/.style={
        circle,
        fill=black,
        inner sep=2pt
    },
    caselabel/.style={
        font=\scriptsize,
        align=left,
        inner sep=1.5pt
    },
    sourcetag/.style={
        font=\tiny,
        text=gray!75!black
    }
]


\draw[axis] (-6.2,0) -- (6.2,0);
\draw[axis] (0,-4.5) -- (0,4.5);

\node[font=\small\bfseries, anchor=north]
    at (-4.8,-0.18) {Internal};

\node[font=\small\bfseries, anchor=north]
    at (4.8,-0.18) {Outsourced};

\node[font=\small\bfseries, align=center, anchor=south]
    at (0,4.65)
    {Issuer-sales instruction documented};

\node[font=\small\bfseries, align=center, anchor=north]
    at (0,-4.65)
    {No issuer-sales instruction documented};


\node[
    font=\footnotesize\itshape,
    text=gray!55,
    anchor=north west,
    align=left
]
at (-5.9,4.0)
{Internal\\+ issuer sales};

\node[
    font=\footnotesize\itshape,
    text=gray!55,
    anchor=north east,
    align=right
]
at (5.9,4.0)
{Outsourced\\+ issuer sales};

\node[
    font=\footnotesize\itshape,
    text=gray!55,
    anchor=south west,
    align=left
]
at (-5.9,-4.35)
{Internal\\+ no instruction documented};

\node[
    font=\footnotesize\itshape,
    text=gray!55,
    anchor=south east,
    align=right
]
at (5.9,-4.35)
{Outsourced\\+ no instruction documented};


\node[casepoint] at (-4.1,-1.35) {};
\node[caselabel,anchor=east]
    at (-4.25,-1.35)
    {Bitcoin / Mt.\ Gox (2011--13)\\
     \textcolor{gray!75!black}{\tiny [Leaked]}};

\node[casepoint] at (-3.5,-2.35) {};
\node[caselabel,anchor=east]
    at (-3.65,-2.35)
    {VZZN (2023)\\
     \textcolor{gray!75!black}{\tiny [DOJ]}};

\node[casepoint] at (-4.0,1.80) {};
\node[caselabel,anchor=east]
    at (-4.15,1.80)
    {TRX (2018--19)\\
     \textcolor{gray!75!black}{\tiny [SEC]}};


\node[casepoint] at (2.0,2.15) {};
\node[caselabel,anchor=west]
    at (2.15,2.15)
    {\textbf{HYDRO (2018--19)}\\
     \textcolor{gray!75!black}{\tiny [DOJ/SEC]}};


\node[casepoint] at (1.15,-0.90) {};
\node[caselabel,anchor=west]
    at (1.30,-0.90)
    {Robo Inu (2022--24)\\
     \textcolor{gray!75!black}{\tiny [DOJ/SEC]}};

\node[casepoint] at (3.55,-1.65) {};
\node[caselabel,anchor=west]
    at (3.70,-1.65)
    {Saitama (2021--23)\\
     \textcolor{gray!75!black}{\tiny [DOJ/SEC]}};

\node[casepoint] at (1.15,-2.45) {};
\node[caselabel,anchor=west]
    at (1.30,-2.45)
    {SaitaRealty (2022--23)\\
     \textcolor{gray!75!black}{\tiny [DOJ/SEC]}};

\node[casepoint] at (3.55,-3.20) {};
\node[caselabel,anchor=west]
    at (3.70,-3.20)
    {NexFundAI--ZM (2024)\\
     \textcolor{gray!75!black}{\tiny [DOJ/SEC]}};

\node[casepoint] at (1.15,-3.65) {};
\node[caselabel,anchor=west]
    at (1.30,-3.65)
    {\textbf{NexFundAI--CLS (2024)}\\
     \textcolor{gray!75!black}{\tiny [DOJ/SEC]}};

\end{tikzpicture}

\caption{Positioning of selected documented cryptocurrency wash-trading cases.
The horizontal axis distinguishes internal manipulation from outside-firm
engagements. The vertical axis records whether an issuer-sales instruction
is documented, rather than whether a net-liquidation quantity is required.
Saitama and SaitaRealty denote the outside-firm engagements in the Appendix.
VZZN denotes the internal promoter scheme; the separate ZM Quant proposal
is recorded in the Appendix. Lower positions indicate that no such instruction
is documented, including for the selected NexFundAI volume plans.
Mt.\ Gox denotes historical account self-matching without a documented
external issuer contract. Dates indicate trading periods. Positions within
categories separate labels visually. \textbf{HYDRO} and
\textbf{NexFundAI--CLS} are the focal cases; bracketed tags identify the
principal evidence source.}

\label{fig:internal_external_contractibility}
\label{fig1}
\end{figure}

\section{Model}
\label{sec:model}

I first review the available documented contractual arrangements for
outsourced wash trading in Supplemental Appendix Table~B1. The most common structure combines a periodic fixed
service fee with trading resources supplied separately by the token issuer.
These resources typically include the issuer's tokens and a counterasset such
as ETH, BTC, or USDT. The issuer also specifies the objectives of the
engagement, either generating trading volume, maintaining liquidity, or
selling issuer inventory. Thus, compensation, trading resources, and trading
objectives are separate components of the arrangement. This structure
motivates the model below and is consistent with models in which
agents are evaluated on multiple dimensions of performance
\citep{felthamPerformanceMeasureCongruity1994,budde_2007}.

I consider a target-based execution problem in which a sales requirement can
be met by changing either net selling or the amount of offsetting trading.
The token issuer hires a specialized market maker for a fixed fee $F$.
Let $\bar V>0$ denote the required level of gross trading activity and
$\bar S\geq0$ a minimum quantity of gross sales. Both targets are expressed
in token units over the assignment period. A volume-oriented benchmark
has no additional sales requirement, $\bar S=0$.\footnote{The gross-sales
constraint is a theoretical specification; HYDRO's documented sales instruction
does not define whether repurchases are deducted. Both model targets use token
units, whereas CLS's operating plan states dollar volume. An absence of a
documented sales requirement does not imply that sales were prohibited.}

Let $B\geq0$ and $S\geq0$ denote the market maker's aggregate purchases and
sales over the assignment period. Gross trading activity is
\begin{equation}
V=B+S,
\label{eq:model_volume}
\end{equation}
and net selling is
\begin{equation}
L=S-B.
\label{eq:model_liquidation}
\end{equation}
A given level of trading activity can therefore be generated with different
combinations of purchases and sales. When purchases and sales are equal,
$L=0$. When sales exceed purchases, $L>0$ and trading reduces the market
maker's token position on net. The distinction between $S$ and $L$ matters:
repurchases can offset gross sales.

The outsourced wash trader, henceforth the market maker, incurs cost
\begin{equation}
C(V,L)=\frac{c_V}{2}V^2+\frac{c_L}{2}L^2,
\qquad c_V,c_L>0.
\label{eq:model_cost}
\end{equation}
The first term captures the costs of generating trading activity. The second
captures the additional cost of net selling. Repeated offsetting trades can
generate substantial volume without requiring the market to absorb a
persistent increase in token supply, whereas net sales must ultimately be
absorbed by other market participants. I use the quadratic specification to
capture increasing marginal execution costs. A price-impact interpretation
is provided in the Appendix.

Conditional on participation and compliance with the targets, the market maker chooses $V$ and
$L$ to minimize
\begin{equation}
C(V,L)
\label{eq:model_problem}
\end{equation}
subject to
\begin{equation}
V\geq\bar V,\qquad
\frac{V+L}{2}\geq\bar S,\qquad
0\leq L\leq V.
\label{eq:model_constraints}
\end{equation}
The sales constraint follows from $S=(V+L)/2$. The last condition follows
from nonnegative purchases and the focus on net selling. Targets are taken
as given performance requirements. The benchmark assumes sufficient trading
capital and token inventory. I consider inventory and market-absorption
constraints in the Appendix.

\medskip
\noindent\textit{Proposition 1.}
Define $r=\min\{1,c_V/c_L\}$. Conditional on feasibility and participation,
the unique least-cost execution is
\begin{equation}
V^*=\max\left\{\bar V,\frac{2\bar S}{1+r}\right\},
\qquad
L^*=\max\{0,2\bar S-V^*\}.
\label{eq:model_solution}
\end{equation}
There are three regimes. If $\bar S\leq\bar V/2$, the volume requirement
already generates sufficient sales through balanced trading:
$V^*=\bar V$ and $L^*=0$. If
$\bar V/2<\bar S\leq(1+r)\bar V/2$, gross activity remains at its target
and additional required sales are met by increasing net selling:
$V^*=\bar V$ and $L^*=2\bar S-\bar V$. Above this threshold, the market
maker expands gross activity, with $L^*/V^*=r$.

\medskip
Purchases and sales satisfy
\begin{equation}
B^*=\frac{V^*-L^*}{2},\qquad
S^*=\frac{V^*+L^*}{2}.
\label{eq:model_BS}
\end{equation}
In the intermediate regime, purchases fall and sales rise by equal amounts
while gross activity remains unchanged. In the high-sales regime, net
selling is costly enough to justify additional turnover when $c_L>c_V$.
Both purchases and sales then increase with the sales requirement. If
$c_L\leq c_V$, the high-sales solution instead consists entirely of sales.

Let $a=\bar S/\bar V$. The directional and two-way balanced shares are
\begin{equation}
D^*=\frac{L^*}{V^*}=\min\{r,\max(0,2a-1)\},
\qquad R^*=1-D^*.
\label{eq:model_shares}
\end{equation}
For $L^*>0$, gross activity per unit of net selling is
\begin{equation}
\frac{V^*}{L^*}=\frac{1}{D^*}.
\label{eq:model_ratio}
\end{equation}
Holding costs and the volume target fixed, a higher sales requirement weakly
increases directional trading until its share reaches $r$. A higher cost of
net selling relative to turnover reduces this limiting share and favors
more offsetting trading. A sales requirement can therefore coexist with
high two-way balance; it need not produce substantial net liquidation.
These predictions are conditional on additional inventory and
market-absorption constraints being slack. The proofs and extensions are
provided in the Appendix.

\section{Data}
\label{sec:data}

\subsection{Enforcement Records}
\label{sec:enforcement}

I first examine publicly available SEC and DOJ enforcement records and
collect blockchain transaction data through publicly available application
programming interfaces (APIs). The enforcement records identify the parties
involved, the period of manipulation, and the underlying trading
arrangements. The blockchain data provide a record of on-chain transactions,
including transaction timestamps, wallet addresses, and token quantities.
Because these sources are publicly recorded and can be independently
verified, the data are comprehensive and reliable. More
importantly, the enforcement records document what the outside market
maker was hired to accomplish, allowing me to identify contractual
objectives independently of the trading patterns that I subsequently
examine.

I begin with the case inventory in Supplemental Appendix Table~B1, drawn from enforcement records and leaked internal records. The inventory distinguishes documented trading cases from related proposed engagements and preserves the evidentiary status of each record. I classify these cases according to whether the manipulation was conducted internally or outsourced and, for outsourced arrangements, according to the trading objectives documented in the underlying records. I then apply a common data requirement to the outsourced cases: the trading objective must be documented independently of the trading outcomes, and purchase and sale quantities must be observable for an explicitly reported trading scope associated with the outside firm's engagement. These requirements leave two arrangements with different documented objectives, NexFundAI--CLS Global and HYDRO--Moonwalkers, for the main comparison.

For the NexFundAI--CLS Global case, the NexFundAI project team contracted CLS Global to generate artificial trading volume for a monthly fee of US\$4,000, payable in Tether. NexFundAI was created at the FBI's direction, and CLS was unaware of the undercover investigation. CLS's trading therefore reflects its execution of an assignment it believed came from a token project. CLS's operating plan estimated US\$32,000 of trading volume per 24 hours using 30 wallets. CLS conducted the wash trading on Uniswap from August 23 to September 18, 2024, providing reconstructed purchases and sales over 27 active days \citep{sec_cls_2024}.\footnote{CLS Global subsequently pleaded guilty in January 2025 and was sentenced on April 2, 2025 \citep{doj_cls_sentenced_2025}.} 

For the HYDRO--Moonwalkers case, Hydrogen Technology Corporation contracted Moonwalkers on October 4, 2018, to provide market-making services and sell the company's HYDRO holdings while limiting price depreciation. Hydrogen agreed to pay US\$9,000 per month in bitcoin. Services began on October 9, and the manipulation continued from October 2018 through April 2019. On October 13, Hydrogen's CEO, Michael Kane, instructed Moonwalkers to sell 200--400 million HYDRO per month. The instruction documents an issuer-sales target without specifying whether repurchases are deducted from that target. HYDRO provides reported company-account purchase and sale quantities for this engagement \citep{sec_hydrogen_2022}.

\subsection{Transaction Data}
\label{sec:transactions}

For NexFundAI, I reconstruct the transactions of the 30-wallet CLS cluster
identified from the enforcement record. The sample covers August 23 through
September 18, 2024 and contains 740 swaps: 374 purchases and 366 sales over
27 active UTC calendar days. These counts match the CLS trading activity
reported by the SEC \citep{sec_cls_2024}. For the analysis, I aggregate the
associated wallets and measure purchases and sales from the perspective of
the controlled cluster.

For HYDRO, account-level exchange transactions are not publicly available.
I therefore use the aggregate purchase and sale quantities reported in the
defense expert declaration filed in the criminal case. The report covers
October 11, 2018 through April 9, 2019 and reports purchases of
3.141 billion HYDRO and sales of 3.618 billion HYDRO.\footnote{United States
v. Kane, No.\ 1:23-cr-20172-KMW, D.E.\ 190-16, physical page 6,
paragraph 11, and physical page 12, paragraph 28. The HYDRO quantities are
source-reported company-account totals rather than independently reconstructed
account-level executions.}
The declaration states that Moonwalkers gained access to Hydrogen's Bittrex
trading account on October 11, 2018.\footnote{D.E.\ 190-16, physical page 8,
paragraph 15.} The reported company quantities therefore provide a trading
scope associated with this engagement. For the interval-bound comparison,
I interpret the reported purchases and sales as completed executions covering
the same set of company accounts over the same reporting period. The bound is conditional on this interpretation and applies to
the reported company totals, rather than an exclusively Moonwalkers trade roster.

The two datasets therefore provide different levels of granularity. NexFundAI
allows me to observe the timing, direction, and wallet-level organization of
trading, while HYDRO provides episode-level purchases and sales. I use the
common information available in both cases to compare gross trading activity,
net selling, and trading composition, and use the finer NexFundAI data for
the additional interval- and wallet-level analyses.

\subsection{Measures}
\label{sec:measures}

I adapt measures of trading activity and order imbalance from
\citet{chordia_order_2002} and their more recent applications in equity and
cryptocurrency markets
\citep{makarov_trading_2020,boehmer_tracking_2021,
brugler_comertonforde_2025,liu_makarov_schoar_2026}.
I separate gross trading activity from net selling because the two capture
different dimensions of the trading arrangement. Gross activity captures how
much trading is generated, while net selling captures the extent to which
sales exceed purchases. This distinction is central to my setting because
volume generation can be achieved through offsetting purchases and sales,
whereas net liquidation requires sales to exceed purchases.

Let $B_t$ and $S_t$ denote token purchases and sales within the reported
trading scope in interval $t$. The common comparison uses episode totals.
For the CLS interval analysis, the primary partition uses UTC calendar days,
with one-hour and six-hour intervals reported as robustness checks.

First, I measure gross trading activity as
\begin{equation}
G_t=B_t+S_t.
\label{eq:Gemp}
\end{equation}
Trading volume is commonly used to measure the amount of trading activity
\citep{chordia_order_2002,brugler_comertonforde_2025}. I measure activity
in token quantities because the contractual objectives in my setting concern
both trading activity and sales of issuer tokens, making token units directly
comparable across purchases and sales within an arrangement. This gross
token-flow measure is distinct from the dollar volume in CLS’s operating plan.

Second, I measure net selling as
\begin{equation}
N_t=S_t-B_t.
\label{eq:Nemp}
\end{equation}
This follows the signed-volume and order-flow literature, which uses the
difference between purchases and sales to measure the direction of trading
\citep{makarov_trading_2020,liu_makarov_schoar_2026}. I orient the measure
so that positive values indicate net selling after purchases offset sales,
measuring the directional outcome of the issuer-sales assignment.

To compare net selling across intervals with different levels of trading
activity, I scale it by gross trading activity:
\begin{equation}
D_t=\frac{S_t-B_t}{S_t+B_t}.
\label{eq:directional}
\end{equation}
Scaling buy--sell imbalance by total trading volume is common in recent
market-microstructure research
\citep{boehmer_tracking_2021,chen_short_interest_2025}.
This normalization allows me to compare the direction of trading across
intervals with substantially different amounts of activity.

Finally, I express the same normalized imbalance as a measure of two-way
balance:
\begin{equation}
R_t=\frac{2\min(B_t,S_t)}{B_t+S_t}=1-|D_t|.
\label{eq:balance}
\end{equation}
Recent work commonly uses absolute order imbalance to measure the degree of
directional trading \citep{brugler_comertonforde_2025}. I express its
complement as two-way balance because my empirical prediction concerns the
extent to which purchases and sales offset one another. Higher values of
$R_t$ therefore indicate more balanced two-way trading. Both $D_t$ and
$R_t$ are defined only when $G_t>0$. Zero-activity intervals are omitted
from unweighted interval averages and contribute zero to gross-weighted totals.

For episode totals $B=\sum_tB_t$, $S=\sum_tS_t$, $G=B+S$, and $N=S-B$,
I report $N/G$, $R_{\mathrm{episode}}=1-|N|/G$, and $G/|N|$. The first
two extend the normalized order-imbalance measures to the full episode.
The final measure expresses gross trading activity per unit of net selling,
which maps directly into the model's distinction between volume generation
and net liquidation.

\section{Findings}
\label{sec:findings}

\subsection{Predicted Trading Composition}
\label{sec:model_findings}

The model separates a requirement for gross sales from net liquidation.
A volume target already generates gross sales when purchases and sales
offset, so an additional sales instruction does not necessarily produce
a directional imbalance. Once the sales requirement binds, the market
maker can satisfy it through more net selling, more gross turnover, or
both. Their relative execution costs determine the composition.

Figure~\ref{fig2} illustrates the three regimes for
$a=\bar S/\bar V$, holding the positive token-unit volume target fixed.
Below $a=1/2$, the sales requirement is met by balanced trading.
Above this point, directional share initially rises while gross activity
remains at its target. Once $a>(1+r)/2$, gross activity expands and
directional share remains at $r=\min\{1,c_V/c_L\}$.
The figure uses illustrative relative-cost values, rather than estimates
for either case. A higher cost of net selling favors more offsetting
turnover and higher balance, even with a substantial gross-sales quota.

The empirical comparison below examines whether the documented issuer-sales
engagement exhibits greater net selling relative to gross activity, as
predicted when the sales requirement binds and additional net-selling
constraints are slack.

\begin{figure}[htbp]
\centering
\begin{minipage}[t]{0.48\textwidth}
\centering
\textit{A. Gross activity relative to its target}\par\smallskip
\begin{tikzpicture}
\begin{axis}[
width=\linewidth,height=0.85\linewidth,
xmin=0,xmax=1.5,ymin=0.9,ymax=2.6,
xtick={0,0.5,1,1.5},ytick={1,1.5,2,2.5},
xlabel={Gross-sales quota / volume target},
ylabel={Optimal gross activity / target},
tick label style={font=\footnotesize},label style={font=\footnotesize},
axis lines=left,ymajorgrids=true,grid style={gray!20},
legend style={at={(0.5,-0.30)},anchor=north,legend columns=3,
font=\footnotesize,draw=none},
]
\addplot[blue!70!black,very thick,mark=none]
  coordinates {(0,1) (0.625,1) (1.5,2.4)};
\addlegendentry{$c_V/c_L=1/4$}
\addplot[orange!85!black,very thick,dashed,mark=none]
  coordinates {(0,1) (0.75,1) (1.5,2)};
\addlegendentry{$c_V/c_L=1/2$}
\addplot[black,thick,densely dotted,mark=none]
  coordinates {(0,1) (1,1) (1.5,1.5)};
\addlegendentry{$c_V/c_L=1$}
\end{axis}
\end{tikzpicture}
\end{minipage}\hfill
\begin{minipage}[t]{0.48\textwidth}
\centering
\textit{B. Two-way balanced share}\par\smallskip
\begin{tikzpicture}
\begin{axis}[
width=\linewidth,height=0.85\linewidth,
xmin=0,xmax=1.5,ymin=0,ymax=105,
xtick={0,0.5,1,1.5},ytick={0,25,50,75,100},
xlabel={Gross-sales quota / volume target},
ylabel={Share of gross activity (\%)},
tick label style={font=\footnotesize},label style={font=\footnotesize},
axis lines=left,ymajorgrids=true,grid style={gray!20},
legend style={at={(0.5,-0.30)},anchor=north,legend columns=3,
font=\footnotesize,draw=none},
]
\addplot[blue!70!black,very thick,mark=none]
  coordinates {(0,100) (0.5,100) (0.625,75) (1.5,75)};
\addlegendentry{$c_V/c_L=1/4$}
\addplot[orange!85!black,very thick,dashed,mark=none]
  coordinates {(0,100) (0.5,100) (0.75,50) (1.5,50)};
\addlegendentry{$c_V/c_L=1/2$}
\addplot[black,thick,densely dotted,mark=none]
  coordinates {(0,100) (0.5,100) (1,0) (1.5,0)};
\addlegendentry{$c_V/c_L=1$}
\end{axis}
\end{tikzpicture}
\end{minipage}
\caption{Execution under a gross-sales requirement.
The horizontal axis is $a=\bar S/\bar V$, holding $\bar V>0$ fixed.
Panel A plots $V^*/\bar V=\max\{1,2a/(1+r)\}$.
Panel B plots $R^*=1-\min\{r,\max(0,2a-1)\}$,
where $r=\min\{1,c_V/c_L\}$.
The relative-cost values are illustrative and not fitted to either case.
Net-selling capacity and working-capital constraints are assumed slack,
and the assignment is assumed feasible and accepted.}
\label{fig2}
\end{figure}

\subsection{Trading under the Two Contractual Objectives}
\label{sec:grossfindings}

I first compare how purchases and sales are combined over the full trading
episode. Because the two arrangements involve different tokens and very
different amounts of trading, the informative comparison is the composition
of each case's own gross activity. I separate gross activity into the portion
that can be matched by purchases and sales and the portion that represents
net directional trading:

\begin{equation}
G=B+S=2\min(B,S)+|S-B|.
\label{eq:decomposition}
\end{equation}

The first component is two-way balanced activity and the second is directional
imbalance. Dividing each component by gross activity puts the two arrangements
on a common scale.

The difference between the two arrangements is large. CLS purchases
469.156 billion NexFundAI tokens and sells 469.795 billion. As a result,
99.932 percent of its gross activity is two-way balanced and net selling
accounts for only 0.068 percent. The cluster generates 1,470.66 units of
gross activity for each unit of net selling.

The reported HYDRO quantities show substantially more directional trading.
Purchases total 3.141 billion HYDRO and sales total 3.618 billion. Net selling
therefore accounts for 7.056 percent of gross activity and the two-way balanced
share is 92.944 percent. HYDRO generates 14.17 units of gross activity for
each unit of net selling.

These patterns are consistent with the contractual objectives. The
volume-oriented arrangement generates almost entirely offsetting turnover,
whereas the arrangement with an issuer-sales instruction directs a much
larger share of trading toward net sales. Relative to gross activity, the
net-selling share in HYDRO is approximately 104 times that of the CLS cluster.

\begin{table}[htbp]
\centering
\caption{Trading Composition under the Documented Objectives}
\label{tab:episode}
\begin{threeparttable}
\small
\begin{tabularx}{\textwidth}{@{}Xrr@{}}
\toprule
Measure & CLS cluster & HYDRO reported company totals\\
\midrule
\multicolumn{3}{l}{\textit{A. Episode trading quantities}}\\
Purchases, $B$ (billion tokens) & 469.156447 & 3.140669\\
Sales, $S$ (billion tokens) & 469.794903 & 3.617556\\
Gross activity, $G$ (billion tokens) & 938.951350 & 6.758225\\
Net selling, $N$ (billion tokens) & 0.638456 & 0.476887\\
\addlinespace
\multicolumn{3}{l}{\textit{B. Trading composition}}\\
Purchases / gross activity (\%) & 49.9660 & 46.4718\\
Sales / gross activity (\%) & 50.0340 & 53.5282\\
Net selling / gross activity (\%) & 0.0680 & 7.0564\\
Two-way balanced share (\%) & 99.9320 & 92.9436\\
Gross activity / absolute net selling & 1,470.66 & 14.17\\
\bottomrule
\end{tabularx}
\begin{tablenotes}[flushleft]
\footnotesize
\item CLS uses reconstructed purchases and sales for the 30-wallet
cluster, August 23--September 18, 2024. HYDRO uses the Hydrogen
company purchase and sale quantities reported in D.E.\ 190-16,
paragraph 28, for October 11, 2018--April 9, 2019, as defined in
Section~\ref{sec:transactions}. Panel A reports quantities in each token's
native units. Panel B divides by each case's gross activity;
$N/G=(S-B)/(S+B)$ and $R=1-|N|/G$.
\end{tablenotes}
\end{threeparttable}
\end{table}

\subsection{Two-Way Balance across Trading Intervals}
\label{sec:dailychoices}

The episode-level comparison could conceal substantial directional trading
over shorter periods. Purchases made early in an episode could offset sales
made much later, making the full episode appear balanced even when trading is
strongly one-sided from hour to hour or day to day. I therefore examine how
closely purchases and sales offset within shorter trading intervals.

For any partition of the episode, gross-weighted two-way balance satisfies

\begin{equation}
R_w=\frac{\sum_{t:G_t>0} G_tR_t}{\sum_tG_t}
=1-\frac{\sum_t|N_t|}{G}
\leq1-\frac{|\sum_tN_t|}{G}
=R_{\mathrm{episode}}.
\label{eq:balance_bound}
\end{equation}

The episode-level balance therefore provides an upper bound on the balance
that can be observed over shorter intervals. For HYDRO, the reported aggregate
purchases and sales conditionally imply an upper bound of 92.9436 percent for every
partition of the same trades.

Figure~\ref{fig3} compares this bound with the trading observed for CLS.
CLS remains highly balanced when the episode is divided into shorter periods.
Gross-weighted balance is 95.90 percent over six-hour intervals and
97.57 percent over daily intervals. Both exceed the conditional upper bound implied
by the reported HYDRO company-account purchases and sales. Shifting the daily boundary by
12 hours produces essentially the same result, with balance of 97.59 percent.

CLS balance rises from 92.35 percent at one hour to 95.90 percent at six
hours and 97.57 percent at one day. This progression shows that offsetting
purchases and sales accumulate over several hours. The near-perfect balance
observed over the full episode is therefore already present over economically
short horizons.

\begin{figure}[htbp]
\centering
\begin{tikzpicture}
\begin{axis}[
width=0.90\textwidth,height=0.46\textwidth,
xmin=0.5,xmax=4.5,ymin=90,ymax=100,
xtick={1,2,3,4},
xticklabels={1 hour,6 hours,24 hours,24 hours: shifted},
ytick={90,92,94,96,98,100},
ylabel={Gross-weighted two-way balance (\%)},
tick label style={font=\small},label style={font=\small},
axis lines=left,ymajorgrids=true,grid style={gray!20},
legend style={at={(0.5,-0.18)},anchor=north,legend columns=2,
font=\footnotesize,draw=none},
]
\addplot[draw=none,fill=gray!15,forget plot] coordinates
{(0.5,90) (4.5,90) (4.5,92.9436076650) (0.5,92.9436076650)} \closedcycle;
\addplot[black,dashed,thick] coordinates
{(0.5,92.9436076650) (4.5,92.9436076650)};
\addlegendentry{HYDRO: conditional upper bound}
\addplot[only marks,mark=*,mark size=3.2pt,blue!70!black] coordinates
{(1,92.3528048348) (2,95.8991670021) (3,97.5698353309) (4,97.5868648697)};
\addlegendentry{Observed CLS balance}
\node[font=\small,anchor=north] at (axis cs:1,92.20) {92.35};
\node[font=\small,anchor=south] at (axis cs:2,96.05) {95.90};
\node[font=\small,anchor=south] at (axis cs:3,97.72) {97.57};
\node[font=\small,anchor=south] at (axis cs:4,97.74) {97.59};
\node[font=\footnotesize,anchor=south west] at (axis cs:0.60,93.00) {HYDRO: 92.94\%};
\end{axis}
\end{tikzpicture}
\caption{Two-way balance across trading intervals. Points report
$R_w=\sum_tG_tR_t/\sum_tG_t$ for the reconstructed CLS cluster.
The dashed line applies equation~(\ref{eq:balance_bound}) to the reported
HYDRO company purchase and sale totals, conditional on the same-scope
execution interpretation in Section~\ref{sec:transactions}.
The shaded area is the portion of the feasible HYDRO range visible within
the 90--100 percent plotting range. The first three partitions use UTC
boundaries; the fourth shifts 24-hour boundaries by 12 hours.
CLS daily and six-hour balance exceed this conditional bound;
the one-hour comparison does not establish an ordering.}
\label{fig3}
\end{figure}

\subsection{Common Control and Sale Replenishment}
\label{sec:distribution}

For CLS, I treat the controlled wallet cluster as the relevant trading actor.
This matters because an outsourced wash trader can distribute purchases and
sales across many controlled wallets. Looking at wallets separately can make
coordinated two-way trading appear directional even when the firm's aggregate
position changes very little.

Figure~\ref{fig4} shows that this distinction is economically large. Across
individual CLS wallets, gross-weighted absolute imbalance is 20.11 percent.
After aggregating the same wallets under common control, absolute imbalance
falls to 0.068 percent. Purchases in some controlled wallets therefore offset
sales made through others.

I next examine how quickly sold quantities are matched to subsequent
purchases. This measures the timing of offsetting execution flows within
the controlled cluster. Let $q_j$ denote the quantity of sale $j$ and
let $m_j(h)$ denote the quantity matched to subsequent purchases within
horizon $h$. The quantity-weighted replenishment rate is

\begin{equation}
Q(h)=\frac{\sum_jm_j(h)}{\sum_jq_j},
\qquad 0\leq m_j(h)\leq q_j.
\label{eq:replenishment}
\end{equation}

Under the seven-day first-in-first-out allocation, CLS cluster purchases
match 76.03 percent of sold quantity within six hours, 98.86 percent within
24 hours, and 99.25 percent within seven days. Restricting matching to the
wallet making the original sale yields 3.55, 16.93, and 66.13 percent.
Allowing purchases across controlled wallets therefore produces substantially
more matched replenishment. Seven days is the maximum follow-up horizon.
Although the CLS trading episode ends on September 18, complete swap-event
coverage of the original NexFundAI pool extends through October 18, 2024,
providing complete seven-day follow-up for all 366 sales within that pool.

Fixed seven-day FIFO assigns purchases to the oldest outstanding sales.
Horizon-specific expiry instead removes sales older than each horizon before
allocating purchases, allowing purchases to match more recent sales. These
rules yield 0.33 versus 49.09 percent at one hour and 76.03 versus 93.85
percent at six hours. At 24 hours they yield 98.86 and 99.25 percent.
The daily replenishment finding is therefore stable across the two rules,
while the shorter-horizon estimates depend substantially on the allocation.

The wallet and replenishment results show how the volume-oriented arrangement
generates its trading activity. Individual wallets take directional positions,
but these trading flows offset across the controlled cluster, and subsequent
purchases match nearly all sold quantity within a day under both cluster rules.
This pattern is consistent with repeated turnover supported by replenishment.

\begin{figure}[htbp]
\centering
\begin{minipage}[t]{0.40\textwidth}
\centering
\textit{A. Wallet and cluster imbalance}\par\smallskip
\begin{tikzpicture}
\begin{axis}[
width=\linewidth,height=1.02\linewidth,
ybar,bar width=30pt,xmin=0.5,xmax=2.5,ymin=0,ymax=24,
xtick={1,2},xticklabels={Individual wallets,CLS cluster},
ytick={0,5,10,15,20},ylabel={Absolute imbalance (\%)},
tick label style={font=\footnotesize},label style={font=\footnotesize},
axis lines=left,ymajorgrids=true,grid style={gray!20},
]
\addplot[bar shift=0pt,fill=orange!65,draw=orange!85!black] coordinates {(1,20.1068739694)};
\addplot[bar shift=0pt,fill=blue!55,draw=blue!70!black] coordinates {(2,0.0679966806)};
\node[font=\footnotesize,anchor=south] at (axis cs:1,20.40) {20.1069\%};
\node[font=\footnotesize,anchor=south] at (axis cs:2,0.65) {0.0680\%};
\end{axis}
\end{tikzpicture}
\end{minipage}\hfill
\begin{minipage}[t]{0.56\textwidth}
\centering
\textit{B. Sold quantity replenished}\par\smallskip
\begin{tikzpicture}
\begin{axis}[
width=\linewidth,height=0.73\linewidth,
xmin=0.8,xmax=5.2,ymin=0,ymax=105,
xtick={1,2,3,4,5},xticklabels={1 hour,6 hours,24 hours,3 days,7 days},
ytick={0,20,40,60,80,100},ylabel={Replenished sold quantity (\%)},
tick label style={font=\footnotesize},label style={font=\footnotesize},
axis lines=left,ymajorgrids=true,grid style={gray!20},
legend style={at={(0.5,-0.18)},anchor=north,legend columns=1,
font=\footnotesize,draw=none},
]
\addplot[blue!70!black,thick,mark=*,mark size=2pt] coordinates
{(1,0.3267715914) (2,76.0314990598) (3,98.8573533490)
 (4,99.2496618826) (5,99.2496618826)};
\addlegendentry{Cluster: seven-day FIFO}
\addplot[orange!85!black,thick,dashed,mark=square*,mark size=2pt] coordinates
{(1,0.1124697204) (2,3.5481426469) (3,16.9260722434)
 (4,41.1468717970) (5,66.1337971860)};
\addlegendentry{Same wallet: seven-day FIFO}
\addplot[black,thick,densely dotted,mark=triangle*,mark size=2pt] coordinates
{(1,49.0893260509) (2,93.8514969115) (3,99.2496618826)
 (4,99.2496618826) (5,99.2496618826)};
\addlegendentry{Cluster: horizon-specific expiry}
\end{axis}
\end{tikzpicture}
\end{minipage}
\caption{Common control and replenishment in the CLS arrangement.
Panel A compares $\sum_a|N_a|/\sum_aG_a$ across individual wallets with
$|\sum_aN_a|/\sum_aG_a$ after aggregating the cluster. Panel B allocates
subsequent purchases to the 366 sales, weighting by sold quantity.
The fixed FIFO rules retain sales for seven days; same-wallet matching
uses purchases by the selling wallet. Horizon-specific expiry removes
sales older than the plotted horizon before matching. Each purchase unit
is used once within each allocation rule. Replenishment measures purchases
of fungible token quantities following sales.}
\label{fig4}
\end{figure}

\section{Conclusion}
\label{sec:conclusion}

I show that trading composition differs across outside-firm assignments
with different contractual objectives. When the assignment is primarily to generate
trading volume, purchases and sales are highly balanced and sold quantities
are rapidly replenished through purchases across controlled accounts. When the
assignment also includes an issuer-sales instruction, trading contains a
substantially larger net-selling component.

The contribution of this paper is to show that outsourced wash trading cannot
be characterized by reported volume alone. Existing research has largely
focused on identifying artificial trading and measuring its prevalence. I show
that gross trading activity and net liquidation are distinct dimensions of
manipulation and that their observed composition differs across contractual
objectives. The same amount of reported activity can
therefore represent extensive recycling of controlled inventory or substantially
more net selling of issuer tokens.

Taken together, these results show that outsourced wash trading varies across
assignments with different documented objectives. Once manipulation is delegated to a specialized outside
firm, contractual objectives and relative execution costs jointly determine
least-cost execution in the model. A binding issuer-sales target can be met through
additional net selling or additional gross turnover. Relative execution costs
govern this choice when inventory and market-absorption constraints are slack. The contract therefore becomes an important part of the
economic organization of manipulation, linking the issuer's objective to the
transaction patterns ultimately observed in the market.

\bibliography{ref}


\clearpage

\section*{Online Appendix}

\subsection*{1. Model Derivations}
\label{app:model}

\subsection*{A. Execution Problem}

The token issuer specifies an assignment $(F,\bar V,\bar S)$, where $F$ is
the fixed service fee, $\bar V>0$ is the required gross trading activity,
and $\bar S\geq0$ is a minimum gross-sales quantity.
Let aggregate purchases and sales be $B$ and $S$. Define
\begin{equation}
V=B+S,\qquad L=S-B.
\label{eq:appendix_flows}
\end{equation}
Because the analysis focuses on net selling, $L\geq0$. Inverting these
relations gives
\begin{equation}
B=\frac{V-L}{2},\qquad S=\frac{V+L}{2}.
\label{eq:appendix_BS}
\end{equation}
Nonnegative purchases require $L\leq V$.

Conditional on accepting the assignment, the market maker solves
\begin{equation}
\min_{V,L}\left\{\frac{c_V}{2}V^2+\frac{c_L}{2}L^2\right\}
\label{eq:appendix_problem}
\end{equation}
subject to
\begin{equation}
V\geq\bar V,\qquad V+L\geq2\bar S,\qquad 0\leq L\leq V.
\label{eq:appendix_constraints}
\end{equation}
Strict convexity gives a unique optimum. The fixed fee affects participation,
but does not affect the least-cost choice conditional on acceptance and
enforcement of the targets.

If $\bar S\leq\bar V/2$, $(V,L)=(\bar V,0)$ satisfies the sales
requirement and minimizes both costly quantities. Otherwise the sales
constraint binds. Substituting $V=2\bar S-L$ reduces the problem to
\begin{equation}
\min_{0\leq L\leq\min\{\bar S,\,2\bar S-\bar V\}}
\left\{\frac{c_V}{2}(2\bar S-L)^2+\frac{c_L}{2}L^2\right\}.
\label{eq:appendix_reduced_problem}
\end{equation}
The unconstrained minimizer is $2c_V\bar S/(c_V+c_L)$. Projecting it onto
the feasible interval and defining $r=\min\{1,c_V/c_L\}$ yields
\begin{equation}
V^*=\max\left\{\bar V,\frac{2\bar S}{1+r}\right\},\qquad
L^*=\max\{0,2\bar S-V^*\}.
\label{eq:appendix_general_solution}
\end{equation}
Equivalently,
\begin{equation}
(V^*,L^*)=
\begin{cases}
(\bar V,0),&0\leq\bar S\leq\bar V/2,\\
(\bar V,2\bar S-\bar V),
  &\bar V/2<\bar S\leq(1+r)\bar V/2,\\
\left(2\bar S/(1+r),\,2r\bar S/(1+r)\right),
  &\bar S>(1+r)\bar V/2.
\end{cases}
\label{eq:appendix_solution}
\end{equation}
In the intermediate regime,
\begin{equation}
B^*=\bar V-\bar S,\qquad S^*=\bar S.
\label{eq:appendix_BSstar}
\end{equation}
Hence $\partial B^*/\partial\bar S=-1$ and
$\partial S^*/\partial\bar S=1$, holding $\bar V$ fixed.
In the high-sales regime,
\begin{equation}
B^*=\frac{1-r}{1+r}\bar S,\qquad S^*=\bar S.
\label{eq:appendix_high_sales}
\end{equation}
Thus additional sales are met with additional purchases when $r<1$,
whereas $r=1$ gives pure selling.

\subsection*{B. Trading Composition}

For $L\geq0$, the gross quantity that can be matched by opposite-direction
trading is $2B=V-L$. The two-way balanced share is
\begin{equation}
R=\frac{2B}{V}=1-\frac{L}{V}.
\label{eq:appendix_balance}
\end{equation}
Let $a=\bar S/\bar V$. At the optimum,
\begin{equation}
D^*=\frac{L^*}{V^*}=\min\{r,\max(0,2a-1)\},
\qquad R^*=1-D^*.
\label{eq:appendix_Dstar}
\end{equation}
Directional share is zero in the balanced regime, increases at rate
$2/\bar V$ with $\bar S$ in the intermediate regime, and is constant
at $r$ in the high-sales regime. Two-way balance has the opposite
comparative statics. For $L^*>0$,
\begin{equation}
\frac{V^*}{L^*}=\frac{1}{D^*},
\label{eq:appendix_ratio}
\end{equation}
which decreases in the intermediate regime and equals $1/r$ in the
high-sales regime. At $L^*=0$ this ratio is undefined.

When $c_L>c_V$, the high-sales solution is
\begin{equation}
V^*=\frac{2c_L}{c_V+c_L}\bar S,\qquad
L^*=\frac{2c_V}{c_V+c_L}\bar S.
\label{eq:appendix_cost_shares}
\end{equation}
Within this regime, a higher $c_L$ increases gross activity and decreases
net selling. More generally, increasing $c_L$ weakly reduces $D^*$ and
weakly increases $R^*$. These results allow gross sales to rise while
directional share remains small. Proposition~1 follows.

\subsection*{C. Why Net Selling Is Costly}

The quadratic net-selling cost can be given a simple market-absorption
interpretation. Suppose the residual price available when placing tokens
with the outside market is
\begin{equation}
p(x)=p_0-\lambda x,\qquad \lambda>0,
\label{eq:appendix_inverse_demand}
\end{equation}
where $x$ is cumulative net quantity sold. Selling $L$ tokens net generates
\begin{equation}
\int_0^Lp(x)\,dx=p_0L-\frac{\lambda}{2}L^2.
\label{eq:appendix_revenue}
\end{equation}
Relative to executing all net sales at $p_0$, the loss is
\begin{equation}
p_0L-\int_0^Lp(x)\,dx=\frac{\lambda}{2}L^2.
\label{eq:appendix_price_impact}
\end{equation}
This motivates $c_L$ as a reduced-form market-absorption cost; the costs of gross
turnover are captured separately by $c_V$. Gross sales and persistent
net supply are different quantities.

\subsection*{D. Price Constraint and Market Absorption}

Some assignments require the market maker to limit price depreciation.
Suppose $p_0\geq\underline p$ and the issuer requires
\begin{equation}
p(L)\geq\underline p.
\label{eq:appendix_price_floor}
\end{equation}
The residual-demand specification implies
\begin{equation}
L\leq D_{\max}\equiv\frac{p_0-\underline p}{\lambda}.
\label{eq:appendix_absorption}
\end{equation}
This is a bound on net selling, not on gross sales. If the benchmark
optimum exceeds $D_{\max}$, the market maker can instead meet its
gross-sales requirement with more offsetting turnover. The price
constraint alone therefore does not make the assignment infeasible when
gross turnover is unrestricted.

For any net-selling cap $K\geq0$, the constrained solution is
\begin{equation}
L_K^*=\min\{K,L^*\},\qquad
V_K^*=\max\{\bar V,2\bar S-L_K^*\}.
\label{eq:appendix_capped_solution}
\end{equation}
To verify this result, for each $L$ the least feasible gross quantity is
$\max\{\bar V,L,2\bar S-L\}$. The resulting cost is convex in $L$,
and capping its unique minimizer at $K$ gives the solution above.
When the cap binds, $L_K^*=K$ and $V_K^*=2\bar S-K$.
Consequently, if $K>0$, further increases in the sales requirement
decrease directional share $K/(2\bar S-K)$. The uncapped monotonic
prediction therefore need not hold under an absorption constraint.

\subsection*{E. Token Inventory}

Let $Q_L$ denote inventory available for terminal net disposition.
This implies $L\leq Q_L$. Combining this terminal inventory restriction
with the price constraint gives
\begin{equation}
K=\min\{Q_L,D_{\max}\}.
\label{eq:appendix_inventory}
\end{equation}
The solution is equation~(\ref{eq:appendix_capped_solution}). If
$K<L^*$, greater available net-selling capacity reduces the turnover
needed to meet the gross-sales requirement. Once $K\geq L^*$, the cap is
slack and further capacity does not change the optimum.

If gross activity also has a technological upper bound $U$, feasibility
requires and is equivalent to
\begin{equation}
\bar V\leq U,\qquad
2\bar S\leq U+\min\{K,U\}.
\label{eq:appendix_feasibility}
\end{equation}
The conditions follow by using the largest feasible gross activity and
net selling. The benchmark assumes adequate
working inventory and counterassets to execute the required turnover.

\subsection*{F. Does the Market Maker Use All Available Tokens?}

Gross trading volume is a flow, whereas token inventory is a stock.
Repeated offsetting trading can therefore generate activity much larger
than the inventory actively used. Define the two-way component
\begin{equation}
A=V-L=2B.
\label{eq:appendix_roundtrip}
\end{equation}
Let $q$ denote token inventory actively used for this component and
$\tau>0$ the gross two-way turnover one unit of active inventory can
support during the assignment. Under this stylized technology,
\begin{equation}
A\leq\tau q,\qquad
q\geq\frac{V-L}{\tau}.
\label{eq:appendix_turnover_technology}
\end{equation}
These are operating-inventory requirements, distinct from the terminal
net-selling bound $Q_L$. In the intermediate sales regime, required
two-way activity is $2(\bar V-\bar S)$ and falls as the sales quota
increases. In the high-sales regime, it equals
$2(1-r)\bar S/(1+r)$ and rises with the quota when $r<1$.
If the net-selling cap binds, it instead equals $2(\bar S-K)$.

If more inventory than this minimum is supplied, the benchmark is
indifferent among feasible deployment choices. It does not predict
that every supplied token is placed in circulation.
A binding operating-inventory constraint would require solving the
execution problem with that additional restriction.

\subsection*{G. Participation}

Let the market maker's outside option be zero. Conditional on feasible
targets and their enforcement, participation requires
\begin{equation}
F\geq C(V^*,L^*).
\label{eq:appendix_participation}
\end{equation}
In the high-sales regime, the minimum fee is
\begin{equation}
F_{\min}
=\frac{2\bar S^2}{(1+r)^2}(c_V+c_Lr^2).
\label{eq:appendix_participation_expanded}
\end{equation}
For an assignment with a binding net-selling cap, the same participation
condition uses $C(V_K^*,L_K^*)$. Additional gross turnover may therefore
restore technological feasibility without making a fixed fee sufficient.
The fixed fee changes willingness to accept the assignment; it does not
alter least-cost execution conditional on acceptance. Targets are given and enforced.
\hfill$\square$

\clearpage
\begin{landscape}
\subsection*{2. Documented Cases and Sample Selection}
\label{app:contracts}

\begingroup
\setstretch{1.0}
\scriptsize
\setlength{\tabcolsep}{3pt}
\renewcommand{\arraystretch}{1.16}
\setlength{\LTleft}{\fill}
\setlength{\LTright}{\fill}
\setlength{\LTcapwidth}{\linewidth}
\captionsetup{font=small,labelfont=bf}
\setcounter{table}{0}
\renewcommand{\thetable}{A\arabic{table}}
\providecommand{\theHtable}{\arabic{table}}
\renewcommand{\theHtable}{case_selection.\arabic{table}}

\noindent
Table~\ref{tab:case_selection} records each case's outsourcing status,
documented contract terms, and inclusion or exclusion from the main trading
comparison. ``Yes'' identifies a documented outside-firm engagement;
proposal-only arrangements are identified separately. The decision column
applies the paper's requirements for an implemented independent-firm
arrangement, documented trading objectives, and available purchase and sale
quantities. Service compensation and trading resources are recorded separately.

\medskip

\begin{longtable}{@{}
>{\raggedright\arraybackslash}p{0.14\linewidth}
>{\raggedright\arraybackslash}p{0.08\linewidth}
>{\raggedright\arraybackslash}p{0.26\linewidth}
>{\raggedright\arraybackslash}p{0.20\linewidth}
>{\raggedright\arraybackslash}p{0.15\linewidth}
>{\raggedright\arraybackslash}p{0.10\linewidth}
@{}}
\caption{Documented Wash-Trading Cases: Outsourcing, Contract Terms and Sample Selection}
\label{tab:case_selection}\\
\toprule
\textbf{Case and period} & \textbf{Documented outsourcing} & \textbf{Contract terms and trading instructions} & \textbf{Fees and resources} & \textbf{Main comparison: use and reason} & \textbf{Primary record}\\
\midrule
\endfirsthead
\multicolumn{6}{l}{\textit{Table \thetable\ continued}}\\[0.08cm]
\toprule
\textbf{Case and period} & \textbf{Documented outsourcing} & \textbf{Contract terms and trading instructions} & \textbf{Fees and resources} & \textbf{Main comparison: use and reason} & \textbf{Primary record}\\
\midrule
\endhead
\midrule
\multicolumn{6}{r}{\textit{Continued on next page}}\\
\endfoot
\bottomrule
\endlastfoot

\textbf{NexFundAI / CLS Global}\newline 2024; trading August 23--September 18. Executed software license and implemented trading plan. &
Yes &
Trading, order-book and spread software. Operating plan estimated US\$32,000 volume per 24 hours using 30 trading wallets.\newline Volume generation was the selected plan. Selling into outside purchases was offered; a fixed issuer-sale requirement is not stated. &
US\$4,000/month in Tether. NexFundAI tokens and ETH supplied August 22; one funding address and 30 trading addresses. Token-tax whitelist requested. &
\textbf{Included.} Implemented volume-generation plan and reconstructed purchases and sales for the 30-wallet cluster. &
\href{https://www.sec.gov/files/litigation/complaints/2024/comp-pr2024-166-cls.pdf}{SEC complaint}\newline Pars.\ 50--60, 65--69.\\[0.12cm]

\textbf{HYDRO / Moonwalkers}\newline Hydrogen Technology Corporation. Signed October 4, 2018; services October 9; renewed engagement through April 2019. &
Yes &
Market making and artificial activity using a trading bot. Initial three-month engagement; trading capital could be replenished using proceeds.\newline Sell company HYDRO while limiting price depreciation. Kane\textquotesingle s October 13 instruction: 200--400 million HYDRO/month, approximately 10 million/day. &
US\$9,000/month, paid in BTC. Initial 5 BTC working capital; HYDRO, BTC and ETH in client exchange accounts. Return 5 BTC plus proceeds net of trading fees. Kane converts BTC proceeds into USD. &
\textbf{Included.} Quantified issuer-sales instruction and reported company-account purchase and sale totals. &
\href{https://www.sec.gov/files/litigation/complaints/2022/comp25553.pdf}{SEC complaint}\newline Pars.\ 103--106, 113--116.\\[0.12cm]

\textbf{NexFundAI / ZM Quant}\newline May 2024. Signed agreement, fee payment, funding and trading documented. &
Yes &
Trading advice and bot-supported project stability; communications specified artificial volume using multiple wallets. Illustration: 100,000 volume through 1,000 transactions at about US\$3 each; no daily period specified.\newline Cash-out offered on sponsor request: distribute tokens, sell small quantities, pool proceeds. A fixed sale quantity is not stated. &
US\$2,500/month in USDT. About US\$5,000 of NexFundAI \emph{plus} US\$5,000 of wETH supplied for production; ETH and tokens separately supplied for testing. &
\textbf{Excluded.} Trading was implemented, but the operator\textquotesingle s trading-wallet roster remains unidentified in the current collection. &
\href{https://www.sec.gov/files/litigation/complaints/2024/comp-pr2024-166-zm-quant.pdf}{SEC complaint}\newline Pars.\ 60--74.\newline \href{https://www.justice.gov/d9/2024-10/zm_quant_et_al._superseding_indictment_0.pdf}{Indictment}\newline Par.\ 49.\\[0.12cm]

\textbf{Saitama / SaitaRealty / ZM Quant}\newline 2021--2023. Paid volume-support engagements. &
Yes &
Artificial volume, continuous charts and exchange maintenance. Expanded quotation covered Saitama on four exchanges and SaitaRealty on one.\newline Issuer-sale instruction: not documented for these engagements. Promoters\textquotesingle\ own sales are separately described. &
Basic fee US\$1,500/month in Tether. Expanded five-service quote: 7,500/month, currency unspecified. Capital request: 5,000--10,000 plus equivalent tokens, denomination unspecified. Client tokens and funded, whitelisted exchange accounts supplied. &
\textbf{Excluded.} Firm-attributed purchase and sale quantities are unavailable; the records do not establish an issuer-liquidation instruction. &
\href{https://www.sec.gov/files/litigation/complaints/2024/comp-pr2024-166-zm-quant.pdf}{SEC complaint}\newline Pars.\ 119--132.\newline \href{https://www.justice.gov/d9/2024-10/zm_quant_et_al._superseding_indictment_0.pdf}{Indictment}\newline Pars.\ 23--35.\\[0.12cm]

\textbf{Robo Inu / Gotbit}\newline Earlier engagement: disclosed fee period February 2022--June 2023; admitted manipulation continued through May 2024. &
Yes &
Volume and price support. Recommended starting volume US\$200,000/day; campaign instructions around US\$1--1.1 million per rolling 24 hours.\newline Mandatory issuer-sale quantity: not documented for this earlier engagement. &
About US\$3,000--6,000/month; payment asset unspecified. Client supplied Robo Inu tokens and USDT for BitMart trading. &
\textbf{Excluded.} Comparable firm-attributed purchase and sale quantities are unavailable for the earlier engagement. &
\href{https://www.sec.gov/files/litigation/complaints/2024/comp-pr2024-166-gotbit.pdf}{SEC complaint}\newline Pars.\ 65--97.\newline \href{https://www.justice.gov/d9/2025-03/usa_v._gotbit_-_plea_agreement.pdf}{Agreed facts}\newline Ex.\ A, par.\ 18.\\[0.12cm]

\textbf{Robo Inu / Gotbit}\newline Renewed agreement signed June 3, 2024. &
Yes &
Meet client-communicated exchange minimum-volume requirements. Fees not due for months requirements are unmet; firm responsible for allocating resources to meet targets.\newline Liquidation priced at 2\% of liquidation generated; 20\% of net trading profit. These are compensation terms, without a fixed sale amount or schedule. &
15,000 USDT upfront for three months on one exchange. Cash-out subaccounts discussed; fresh trading-capital allocation not stated in quoted terms. &
\textbf{Excluded.} Comparable purchase and sale coverage is unavailable. The liquidation fee does not specify a required sale quantity. &
\href{https://www.justice.gov/d9/2024-10/andriunin_complaint_0.pdf}{FBI affidavit}\newline Pars.\ 44--48.\newline \href{https://www.sec.gov/files/litigation/complaints/2024/comp-pr2024-166-gotbit.pdf}{SEC complaint}\newline Pars.\ 124--130.\\[0.12cm]

\textbf{Saitama / Gotbit}\newline 2023. LBank and XT volume support implemented. &
Yes &
LBank instruction to reduce volume below 200,000, denomination unspecified; later maintain at least 10,000 USDT/day. Fee-exempt accounts requested.\newline Delegated issuer-sale requirement: not documented. Promoters\textquotesingle\ sales are not assigned to Gotbit by these terms. &
Two approximately US\$5,000-valued cryptocurrency payments, March 17 and April 19; transferred asset unspecified. Low/zero-fee exchange accounts obtained. &
\textbf{Excluded.} Firm-attributed purchase and sale quantities are unavailable; promoter sales do not establish delegated issuer liquidation. &
\href{https://www.justice.gov/d9/2024-10/andriunin_complaint_0.pdf}{FBI affidavit}\newline Pars.\ 55--64.\newline \href{https://www.justice.gov/d9/2025-03/usa_v._gotbit_-_plea_agreement.pdf}{Agreed facts}\newline Ex.\ A, par.\ 19.\\[0.12cm]

\textbf{Lexobit / Gotbit}\newline Separate undercover engagement. Agreement May 22, 2024; trading August 2024. &
Yes &
Trading-volume system with client-directed changes to charts and activity. Firm supplied a roster of 36 trading wallets.\newline Contract prices liquidation at 2\% and net profit at 20\%. A fixed token-sale requirement is not stated. &
US\$15,000 for three months, paid to a cryptocurrency wallet; coin unspecified. Requested 5 ETH plus 4\% of supply. Delivery of 5 ETH documented. &
\textbf{Excluded.} An attributed trade history has not been recovered. The liquidation commission is not a quantified sale requirement. &
\href{https://storage.courtlistener.com/recap/gov.uscourts.cand.446982/gov.uscourts.cand.446982.1.0.pdf}{Tsao indictment}\newline Pars.\ 21--32, especially 25--27 and 30--32.\\[0.12cm]

\textbf{Powerlink / Vortex}\newline Separate undercover engagement. Agreement September 30, 2024; trading October 9. &
Yes &
Formal agreement prohibited unlawful manipulation. Operating instructions accepted US\$15,000 first-hour volume and later price target US\$0.0013, described as a 30\% increase, with 0.7 ETH purchases authorized.\newline Contract: 1.5\% of monthly cash-out and 15\% of monthly net profit. Client instructed cash-out one week after launch. &
US\$3,900/month; fee payment asset unspecified. Supplied approximately US\$8,500 of ETH and 50 million PLNK. ETH-funded token purchases expressly authorized. &
\textbf{Excluded.} Comparable purchase and sale coverage is unavailable; planned later cash-out is not established in the recovered episode. &
\href{https://storage.courtlistener.com/recap/gov.uscourts.cand.455428/gov.uscourts.cand.455428.1.0.pdf}{Gora indictment}\newline Pars.\ 29, 35--52.\\[0.12cm]

\textbf{Powerlink / Contrarian and Antier personnel}\newline Separately described undercover engagement; trading October 8--9, 2024. &
Yes\newline Contrarian &
Volume, liquidity and price support. Funding request described a prospective 200--250\% price increase; no fixed wash-volume quota stated.\newline Liquidation strategy described in filed memoranda. Quantified sale requirement and percentage compensation: not stated. &
General service fee about US\$3,000/month; US\$9,000 retainer charged. Requested 50 million tokens and US\$30,000 more of ETH. Payment coin and delivery of requested capital unspecified. &
\textbf{Excluded.} Attributed purchase and sale quantities are unavailable. Antier\textquotesingle s separate contractual role is not established. &
\href{https://storage.courtlistener.com/recap/gov.uscourts.cand.455832/gov.uscourts.cand.455832.44.0.pdf}{Singh memorandum}\newline D.E.\ 44, pp.\ 1--4.\newline \href{https://storage.courtlistener.com/recap/gov.uscourts.cand.455835/gov.uscourts.cand.455835.25.0.pdf}{Sharma memorandum}\newline D.E.\ 25, pp.\ 3--4.\\[0.12cm]

\textbf{VZZN / ZM Quant}\newline March 2023 quotation. Proposed engagement. &
Proposal only &
Offered DEX volume creation and a continuous-chart trading bot. Acceptance and implementation of this proposal are not established in the cited passages.\newline No agreed issuer-sale quantity stated in the quotation. &
Quoted 1,500 USDT for DEX volume creation; 1,500 USDT/month/pair for chart and volume bot. Delivered trading resources: not documented. &
\textbf{Excluded.} Acceptance and implementation of the quoted engagement are not established. &
\href{https://www.justice.gov/d9/2024-10/zm_quant_et_al._superseding_indictment_0.pdf}{Indictment}\newline Pars.\ 36--38.\\[0.12cm]

\textbf{NexFundAI / MyTrade}\newline August--September 2024 discussions. Services agreed in discussion; proposed execution. &
Yes\newline Services agreed &
Software allowed client-set volume ranges. Referrer pitched 100,000--200,000 additional daily volume. Completed NexFundAI trading not established by the cited passages.\newline Cash-out tools and founder-sale advice offered; no assigned sale amount stated. &
Referrer quoted US\$500/month software; Zhou offered US\$2,000/month advice and requested US\$2,000 upfront. General tool used client tokens and USDT; NexFundAI funding and payment asset not specified. &
\textbf{Excluded.} Completed NexFundAI trading and an attributed purchase and sale history are not established. &
\href{https://www.justice.gov/d9/2024-10/zhou_information_0.pdf}{Zhou information}\newline Pars.\ 17, 25--34.\\[0.12cm]

\textbf{Bitcoin / Mt.\ Gox}\newline 2011--2013 account sample. Historical self-matching records. &
No documented external contract &
Wash trades identified when the same trader ID appears on both sides. The records measure account trading rather than an outsourced issuer agreement.\newline Issuer-liquidation mandate: not applicable to the observed account identification. &
BTC/fiat trade quantities observed. External service fee, sponsor-supplied inventory and contractual conversion obligations not recorded. &
\textbf{Excluded.} Historical account self-matching records do not document a comparable issuer--independent-firm arrangement. &
\href{https://pubsonline.informs.org/doi/10.1287/mnsc.2021.01448}{Aloosh--Li}\newline \emph{Direct Evidence of Bitcoin Wash Trading}.\\[0.12cm]

\textbf{Coinbase / Employee A}\newline August--September 2016. Internal employee trading; separate automated programs operated during 2015--2018. &
No\newline Internal &
Employee self-trading in LTC/BTC sought apparent interest and liquidity. The order separately describes trades between Hedger and Replicator.\newline Outside issuer-liquidation instruction: not documented. &
Internal exchange trading. No independent issuer--firm fee or inventory grant identified. US\$6.5 million was a regulatory penalty. &
\textbf{Excluded.} Internal employee and exchange activity, rather than an independent issuer--firm engagement. &
\href{https://www.cftc.gov/PressRoom/PressReleases/8369-21}{CFTC order}\newline Docket 21-03, II.A and II.C, pp.\ 2--4.\\[0.12cm]

\textbf{TRX / Sun, Tron and Rainberry}\newline Approximately February 2018--February 2019. Internal scheme described in SEC allegations. &
No\newline Internal &
Raise volume and support price to facilitate TRX sales. Directed 2.5 BTC/day budget across exchanges; this is a trading budget.\newline Internal sponsor-sales objective; independent outside-firm sale requirement not documented. &
US\$10,000 bank wire and Sun\textquotesingle s TRX described. Trading pairs included TRX/USD, TRX/USDT and TRX/BTC. External retainer not stated. &
\textbf{Excluded.} Issuer-directed internal trading, rather than a documented independent-firm engagement. &
\href{https://www.sec.gov/files/litigation/complaints/2026/comp26496.pdf}{Amended SEC complaint}\newline Pars.\ 162, 166--182.\\[0.12cm]

\textbf{Binance.US / BAM / Sigma Chain}\newline Alleged activity September 2019--June 2022. Affiliated market maker; SEC case dismissed in May 2025. &
No\newline Affiliated &
Management requested consistent or higher trading volume from the affiliated maker. Reported volume shares are outcomes.\newline Independent token-issuer liquidation agreement or quantified issuer-sale instruction: not documented. &
Independent issuer retainer and token/counterasset allocation not identified in the wash-trading passages. &
\textbf{Excluded.} The market maker was affiliated with the exchange rather than an independent issuer contractor. &
\href{https://www.sec.gov/files/litigation/complaints/2023/comp-pr2023-101.pdf}{SEC complaint}\newline Pars.\ 185--192, 268--275.\newline \href{https://www.sec.gov/enforcement-litigation/litigation-releases/lr-26316}{Dismissal}\newline LR 26316.\\[0.12cm]

\textbf{VZZN / Thompson and collaborators}\newline January--December 2023 alleged promoter scheme. &
No\newline Promoter organized &
Apparent volume and price inflation to induce outside purchases. Coordinated buying and multiwallet pump discussions documented.\newline Promoters intended to sell their controlled VZZN at a profit. This is separate from ZM Quant\textquotesingle s service quotation. &
Internal trading. No accepted independent-firm fee, inventory allocation or quantified delegated sale requirement identified. &
\textbf{Excluded.} Promoter-organized trading, distinct from ZM Quant\textquotesingle s proposed external engagement. &
\href{https://www.justice.gov/d9/2024-10/thompson_information_0.pdf}{Thompson information}\newline Pars.\ 4--7, especially 6(c)--(d) and 7(h), (j).\\[0.12cm]

\textbf{Saitama / internal promoters}\newline July 3--4, 2021. Related internal trading instructions. &
No\newline Promoter organized &
Coordinated purchases of 2--3 trillion tokens per transaction and rotation across 2--3 wallets to create a decentralized appearance.\newline A participant stated an intention to sell after raising price. These instructions concern promoter trading. &
No independent trading-firm retainer or delegated inventory grant stated. Coordinated-buy instructions are recorded separately from self-matching wash trades. &
\textbf{Excluded.} Internal coordinated trading rather than delegated execution by an independent firm. &
\href{https://www.sec.gov/files/litigation/complaints/2024/comp-pr2024-166-saitama.pdf}{SEC complaint}\newline Pars.\ 74, 78, 80, 82--84.\\[0.12cm]
\end{longtable}

\noindent Targets, proposals, and realized trading are distinguished as reported in
the source. ``Not documented'' means that the cited record does not establish
the relevant term. A liquidation commission is distinguished from an
instruction or requirement to sell issuer holdings. Complaints and
indictments report government allegations; cited plea materials contain
agreed facts. The VZZN promoter scheme and ZM Quant's proposed VZZN
engagement are treated as separate cases.

\endgroup
\end{landscape}
\clearpage

\end{document}